\documentclass[aps,twocolumn,nofootinbib,superscriptaddress]{revtex4}
\usepackage{graphicx}
\usepackage{adjustbox}
\usepackage{dcolumn}
\usepackage{bm}
\usepackage{multirow}
\usepackage{amsmath}
\usepackage[letterpaper, hmargin = {2cm}, vmargin = {2cm}]{geometry}
\usepackage{tablefootnote}

\usepackage{graphicx,bm}
\usepackage{bm,color}

\newcommand{\bL}{\begin{Large}}
\newcommand{\eL}{\end{Large}}
\newcommand{\be}{\begin{equation}}  
\newcommand{\ee}{\end{equation}}
\newcommand{\ba}{\begin{eqnarray*}}
\newcommand{\ea}{\end{eqnarray*}}

\begin{document}

\title{Isospin symmetry breaking in the mirror pair $^{73}$Sr-$^{73}$Br}
\author{ S.~M.~Lenzi},
\affiliation{Dipartimento di Fisica e Astronomia, Universit\`a degli Studi di Padova, 
\mbox {and INFN, Sezione di Padova, I-35131 Padova, Italy  }}

\author{A.~Poves} 
\affiliation{Departamento de F\'isica Te\'orica  and IFT-UAM/CSIC, \mbox{Universidad
 Aut\'onoma de Madrid, 28049 Madrid, Spain}}

\author{A.~O.~Macchiavelli}
\affiliation{Nuclear Science Division, Lawrence Berkeley National Laboratory, Berkeley, California 94720, USA}

\date{\today}

\begin{abstract}
 The recent experimental observation of isospin symmetry breaking (ISB) in the ground states  of the $T=3/2$ mirror pair $^{73}$Sr - $^{73}$Br is theoretically studied 
 using large-scale shell model calculations.  The large valence space and the successful
PFSDG-U effective interaction used for the nuclear part of the problem capture possible structural changes and provide a robust basis to treat the ISB effects of both electromagnetic and non-electromagnetic origin.  The calculated shifts and mirror-energy-differences are consistent with the 
 inversion of the $I^{\pi}$= 1/2$^{-}, 5/2^{-}$ states between $^{73}$Sr - $^{73}$Br,
 and suggest that the role played by the Coulomb interaction is dominant.  An isospin breaking contribution of nuclear origin is estimated to be $\approx 25$~keV.  

 \end{abstract}

\maketitle

\section{Introduction}
In a recent article entitled {\sl ``Mirror-symmetry violation in bound nuclear ground states"}~\cite{Hoff20}, Hoff and collaborators reported the results of an experiment carried out at the National Superconducting Cyclotron Laboratory (NSCL), in which the decay of the proton-rich,
$T=3/2$, $T_{z}$=-3/2, isotope $^{73}$Sr was studied. Following a detailed and convincing  analysis of the experimental data they conclude that its ground state has $I^{\pi}$= 5/2$^{-}$. This observation is at odds with its mirror $T=3/2$, $T_{z}$=3/2 partner $^{73}$Br which has a $I^{\pi}$= 1/2$^{-}$ground state, and thus the topic of their work.
The theoretical interpretation, which accompanies the
 paper, cannot reproduce the inversion and the authors conclude with two main points, one related to the well known Thomas-Ehrman shift~\cite{TE1,TE2}: (sic) {\it Such a mechanism is not immediately apparent in the case of $^{73}$Sr / $^{73}$Br, and it may be that charge-symmetry-breaking forces need to be incorporated into the nuclear Hamiltonian to fully describe the presented results}, and the other one related  to possible structural effects: (sic) {\it (the) inversion could be due to small changes in the two competing shapes, particularly their degree of triaxiality, and the coupling to the proton continuum in the (Isobaric Analogue State) IAS of $^{73}$Rb}. 

Besides the fact that the $I^{\pi}$= 1/2$^{-}$ in $^{73}$Sr is an excited state, there is no information available about its location. On the contrary, the level scheme of 
$^{73}$Br is better known with an $I^{\pi}$= 5/2$^{-}$ state at 27~keV,   an $I^{\pi}$= 3/2$^{-}$ at 178~keV and another 
$I^{\pi}$= (3/2$^{-}$, 5/2$^{-}$) state at 241~keV. Given the above it seems opportune to comment already that the Mirror Energy Difference (MED) of the  1/2$^{-}$ arising from 
the mirror symmetry violation can be as low as $\sim$30~keV. Notice that MED's as large as 300~keV have been measured for  the  2$^+$ states of the
$^{36}$Ca-$^{36}$S mirror pair, which can be understood without invoking threshold effects \cite{doornenbal}. Even further, in the same mirror
pair, a prediction of a huge MED of 700~keV for the first excited 0$^+$ states has been made in Ref.~\cite{valiente}, again without the need
of threshold effects.  
There is abundant experimental and theoretical work on the subject of the MED's  
which we believe provides a natural framework to interpret the new data.    Actually, Ref.~\cite{henderson20} places the new result within the context of the extensive body of available data  and the authors concluded that, being entirely consistent with normal behavior, the inversion  does not provide  further insight into isospin symmetry breaking (ISB).

Here, in line with the findings of Refs.~\cite{Zuker02,Bentley07}, we propose an 
explanation based on the configuration interaction shell model (SM-CI) to treat the nuclear (isospin conserving)  part  of the problem, plus a detailed analysis of both Coulomb and other ISB effects. The large valence space and the well established effective interaction we use allow us to describe deformed nuclei in the laboratory frame without the restriction to axially symmetric shapes as considered in Ref.~\cite{Hoff20}.

\section{The Shell Model framework}

 \subsection{The nuclear input}
 We describe the $A=73$, $T=3/2$ system with the isospin conserving  effective interaction 
 PFSDG-U~\cite{Nowacki16} which has been successful for a large region of nuclei, from the $pf$-shell to the $N=40$ and $N=50$ islands of inversion. Recently applied to the structure of $^{78}$Ni~\cite{Ryo19},  it can be considered as an extension of the LNPS interaction~\cite{Lenzi10} which encompasses nuclei at and beyond $N=50$.

\begin{table}[h]
\caption{Valence space and single particle energies used in the present  SM-CI calculations.} 
\bigskip
\begin{tabular*} {\linewidth}{@{\extracolsep{\fill}}|c|cccccc|}
\hline
   Orbit&  ${0f_{7/2}}$ & ${1p_{3/2}}$ & ${0f_{5/2}}$ &   ${1p_{1/2}}$ & ${0g_{9/2}}$ & ${1d_{5/2}}$  \\
 \hline
SPE (MeV) &-8.363&-5.93& -1.525& -4.184& -0.013& 0.937\\
  \hline
\end{tabular*}
\label{spes}
\end{table}

  The PFSDG-U interaction,  defined for the full $pf$+$sdg$ shells, is here used in the valence space given by the orbits:
   ${0f_{7/2}}$, ${1p_{3/2}}$, ${0f_{5/2}}$,   ${1p_{1/2}}$, ${0g_{9/2}}$ and ${1d_{5/2}}$ 
  with  the single particle energies (SPE) taken directly from the experimental spectra of 
   $^{41}$Ca as summarized in Table~\ref{spes}. In the present calculation an inert core of  $^{56}$Ni is adopted and the number of excitations across $N=Z=40$
   are limited to four, to achieve convergence for the states of interest which have  \mbox{dimension $\approx$ 10$^9$.}
    The isospin conserving (nuclear only)  calculation  produces a ground state $I^{\pi}$= 5/2$^{-}$ and the first excited state,  $I^{\pi}$= 1/2$^{-}$ 
  at  21~keV as shown schematically in Figure~\ref{fig1} and in agreement with the new measurement for $^{73}$Sr. A $I^{\pi}$= 3/2$^{-}$ is found at 288~keV. With this as our starting
  point, we shall next turn our attention to the role of the different ISB effects, responsible for the inversion of states in $^{73}$Br.

\subsection{Isospin symmetry breaking analysis}

\begin{figure}
\begin{center}
\includegraphics[width=8cm]{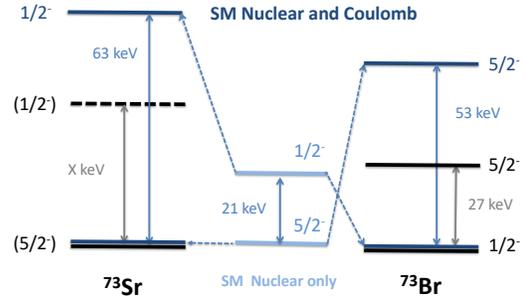}
\caption{(Color online) Shell model results (light-blue and blue levels), compared to the experimental data (black and dashed levels). The SM isospin conserving result with only the nuclear interaction is shown in the middle of the panel.}
\label{fig1}

\end{center}
\end{figure}

  In the following we consider two methods to account for the ISB effects. \\
\\
\noindent
{\bf Method 1:}
 The  Coulomb interaction, $V_C$, is anticipated to be the most important mechanism contributing to the isospin breaking.  In this first approach, it is simply added to the nuclear one in the SM-CI calculation: 
 
 \begin{equation}
     H = H_{N} + V_{C}
\end{equation} 
 We have verified that non-perturbative and perturbative  treatments give almost identical  results.  In the former, the Hamiltonian in Eq. 1 is directly diagonalized for each of the two mirror nuclei:
 
\begin{equation}
 H | ^{73}Sr, I^{\pi} \rangle = E_{I^\pi}(^{73}Sr) |^{73}Sr,I^{\pi} \rangle  \nonumber \\   
 \end{equation}
 \begin{equation}
H |^{73}Br,I^{\pi} \rangle = E_{I^\pi}(^{73}Br) |^{73}Br,I^{\pi} \rangle  \nonumber
 \end{equation} 
\noindent
In the latter, the eigenstates of $H_{N}$, $|A,T, I^{\pi}\rangle$, 
are used to compute the expectation value of the Coulomb interaction for each nucleus:
\begin{equation}
     \delta E_{pert}(^{73}Sr,I^{\pi}) = \langle 73,3/2,I^{\pi}|V_{C}(^{73}Sr)|
     73,3/2,I^{\pi}\rangle \nonumber
\end{equation}  
\begin{equation}
     \delta E_{pert}(^{73}Br, I^{\pi}) = 
     \langle 73,3/2,I^{\pi}|V_{C}(^{73}Br)|
     73,3/2,I^{\pi}\rangle \nonumber
\end{equation}  
\noindent
 \noindent
  $V_C$ can be divided into three terms: core, one-body and two-body.
  With the indexes $m,n$ representing the protons in the core and $i,j$ the valence protons, we have
 \begin{equation}
 V_{C,Core}=\sum_{n,m}  e^2/r_{n,m}\nonumber
\end{equation}
 \begin{equation}
  V_{C,1B}=  \sum_j n_j (\sum_n e^2/r_{n,j}) \nonumber
\end{equation}
\begin{equation}
     V_{C,2B}    =  \sum_{i,j}  e^2/r_{i,j} \nonumber
\end{equation}
 The first term is the same for both nuclei and is not considered further. The one-body
 term affects only the single particle energies of the proton orbits. We adopt the experimental spectrum of $^{41}$Sc, where a lowering of 225~keV of the energies of the $p$ orbits relative to the $f$ orbits is observed. The two-body Coulomb matrix elements are calculated
   with harmonic oscillator (HO) wave functions using 
   $\hbar \omega = 45 A^{-1/3} -25 A^{-2/3}$~ MeV. Their expectation values
   are denoted by $C1$ and $C2$ respectively.  The results for the two mirror isotopes, including the individual contributions, are given in Table \ref{coul-I} and illustrated in Figure~\ref{fig1}.

 \begin{table}[h]
\caption{Method 1. Isospin symmetry breaking contributions to the excitation energies of the lower states in $^{73}$Sr  and $^{73}$Br, C1 (1B) and C2 (2B) (in keV). They are added to the nuclear only values to produce the Total and MED columns, the ones to be eventually compared with experiment.} 
\bigskip
\begin{tabular*} {\linewidth}{@{\extracolsep{\fill}}|c|c|ccc|ccc|c|}
\hline
 $I^{\pi}$&  Nuclear && $^{73}$Sr&&& $^{73}$Br&&MED\\
 & & $C1$  & $C2$ &  Total &
   $C1$  & $C2$  & Total &\\
\hline  
 $5/2^-$  & 0 & 0&0&  0 &53 & 0&  53& 0\\
 $1/2^-$ & 21& 25&17&63 & -27&6& 0&116 \\
 $3/2^-$& 288& 3& -79&212 & 55&18 &361& -104\\
\hline
\end{tabular*}
\label{coul-I}
\end{table}

It is clearly seen that: i) this approach produces the desired inversion in  $^{73}$Br, and, ii) it is the one-body part of the Coulomb interaction, 
{\it i.e.} the shift in the proton single particle energies of the $p-$orbits relative to the $f-$orbits, which is responsible for this phenomenon. If we
 shift the proton SPE's of the $g-$ and $d-$orbits by the same quantity than the $p-$orbits we obtain
   qualitatively the same results, and do not change appreciably even if we double the SPE correction.  It is important to note that the difference of SPE between protons and neutrons, taken from the experimental data, may not be only of electromagnetic origin.  
   
  The MED's are defined as the difference between the excitation energy of analogue states, thus putting the MED for the ground states to zero~\cite{Zuker02}, which have in general the same spin and parity. As this is not the case here, we calculate the MED with respect to the $5/2^-$ state, that is the lowest state for the pure nuclear field. We report in the last column of Table~\ref{coul-I} the MED obtained as
   \begin{equation}
     \mathrm{MED}_{I^\pi}=E^*_{I^\pi}(^{73}{\rm Sr})-E^*_{I^\pi}(^{73}{\rm Br})\nonumber
\end{equation}  
   
\noindent
where $E^*_{I^\pi}= E_{I^\pi}-E_{5/2^-}$. \\

\noindent
{\bf Method 2:}
  Here we follow the  approach discussed in the review article~\cite{Bentley07} that considers several contributions to the MED:

\medskip
\noindent
 {\sl Multipole Coulomb C$_M$}. It is constructed as the Coulomb 2B in Method 1, the only difference is that only the multipole part of the two-body Coulomb matrix elements is considered. It is sensitive to microscopic features such as the change of single-particle spin recoupling and alignment.
 
\medskip
\noindent
 {\sl Single-particle energy corrections $C_{\ell s}$ and  $C_{\ell \ell}$}. Starting from identical single-particle orbits for protons and neutrons, given in Table~\ref{spes}, relative shifts due to the electromagnetic spin-orbit interaction $C_{\ell s}$~\cite{Nolen69} 
and the orbit-orbit term $C_{\ell\ell}$~\cite{Duflo02} are introduced. \\

\noindent
The electromagnetic spin-orbit interaction is:
 \begin{equation}
 V_{\ell s} = (g_s-g_\ell) \frac{1}{2 m_N^2 c^2}\left (-\frac{1}{r}\frac{dV_C}{dr}\right )~\vec{\ell}.\vec{s} \nonumber 
  \end{equation}
 where $g_l$ and $g_s$ are the g-factors and $m_N$ the nucleon mass. The correction is given by
  \begin{equation}
 C_{\ell s} \simeq 14.7 (g_s-g_\ell)(\frac{Z}{A})[ \ell(\ell+1)+ s(s+1)-j(j+1)] ~\mathrm{keV} \nonumber
  \end{equation}
 which, although $\sim$50 times smaller than the nuclear spin-orbit interaction, its effect on the excitation energies can be of several tens to hundreds of keV.
\noindent
 It is clear that this interaction contributes differently on protons and neutrons. \\

\noindent
The $C_{\ell\ell}$ energy correction has been deduced in Ref.~\cite{Duflo02} and is given by
 \begin{equation}
 C_{\ell \ell} = \frac {-4.5 Z_{cs}^{13/12} [2\ell(\ell+1)-N(N+3)]}{A^{1/3}(N+3/2)} ~\mathrm{keV}\nonumber
 \end{equation}

\noindent
with $Z_{cs}$ the atomic number of the closed shell. For  $A=73$, $Z_{cs}$=20 and the corresponding HO principal quantum numbers $N=3$ and $N=4$, the energy shifts to be added to the bare energies in Table~\ref{spes} are reported in Table \ref{ll-ls}.

\begin{table}[h]
\caption{Method 2. Energy corrections introduced by the electromagnetic $C_{\ell s}$ and  $C_{\ell\ell}$ terms to the SPE's of neutrons and protons (in keV) }
\bigskip
\begin{tabular*} {\linewidth}{@{\extracolsep{\fill}}|c|cccccc|}
\hline
                &  ${0f_{7/2}}$ & ${1p_{3/2}}$ & ${0f_{5/2}}$ &   ${1p_{1/2}}$ & ${0g_{9/2}}$ & ${1d_{5/2}}$  \\
 \hline
  neutrons ($\ell s$) & 52.5& 17.5& -70&-35 &70  & 35\\
 protons ($\ell s$+$\ell\ell$) 
   &-100 &  65 & 47& 128& -144 & 38\\ 
  \hline
\end{tabular*}
\label{ll-ls}
\end{table}    

The corrections of electromagnetic origin introduced so far  have no free parameters and affect the excitation energy of the analogue states in each of the mirror nuclei. 
In the following we discuss two additional corrections purely of  isovector character. Therefore we know their contributions to the MED's, but ignore their effect on each mirror partner separately. Both terms are empirical and schematic.  

\medskip
\noindent
  {\sl Radial term C$_r$}. Of Coulomb origin, it takes into account changes of the nuclear radius for each excited state. These changes are due to differences in the nuclear configuration that depend on the occupation number of the orbits. Low-$\ell$ orbits have larger radius than the high-$\ell$ orbits in a main shell. 
This has a sizable effect in the MED: Protons in larger orbits suffer less repulsion than those in smaller orbits, which reflects in the binding energy of the nuclear states. Originally introduced in~\cite{Lenzi01}, the halo character of low-$\ell$ orbits has been recently discussed in detail in~\cite{BLZ}. The isovector polarization effect in mirror nuclei tends to equalize proton and neutron radii. Thus, the contribution of the radial term to the MED at spin $I^\pi$ can be parametrized as a function of the average of proton and neutron radii, considering the change in the occupation of low-$\ell$ orbits between the ground state (gs) and the state of angular momentum $I^\pi$~\cite{Bentley07}:
\begin{equation}
 C_r(I^\pi)= 2 |T_z|  \alpha_r  \left(\frac{n_{\pi}(gs)+n_{\nu}(gs)}{2} - \frac{n_{\pi}(I^\pi)+n_{\nu}(I^\pi)}{2}\right).\nonumber
\end{equation}
\noindent

The value 
$\alpha_r$=200~keV, has been used in extensive studies of MED's in the $pf$-shell~\cite{Bentley07}.
In the present case, since we are also filling the shell $g_{9/2}$ and $d_{5/2}$ orbits, we have to include them as they have larger radii than the $f$ orbits as well. We adopt the same value $\alpha_r=200$ keV for the $p_{1/2}$ orbit, 
$\alpha_r=100$ keV for the $p_{3/2}$ orbit that is almost full~\cite{BZ}, and a larger value of $\alpha_r=300$ keV for the $N=4$~ $g_{9/2}$ and $d_{5/2}$ orbits. The estimated radial contribution is $C_r(1/2^-)=-16$ keV.

\medskip
\noindent
{\sl Isospin--symmetry breaking interaction V$_{B}$}. This is an isovector correction deduced from the $A=42$, $T=1$ mirrors in Ref.~\cite{Zuker02} and more recently modified and generalized in Ref.~\cite{Bentley15}. It consists of a difference of -100 keV between the $I=0, T=1$ proton-proton and neutron-neutron matrix elements. Originally introduced for the $f_{7/2}$ shell, here we apply it to all orbitals in the model space.
  
\medskip
\noindent 
Taking into account all the corrections above we compute the MED's for the $^{73}$Sr and $^{73}$Br mirror pair in first order perturbation theory as,
\begin{eqnarray}
    \mathrm{MED}_{I^\pi} & = & E_{I^\pi}^*(^{73}Sr)-E_{I^\pi}^*(^{73}Br) \nonumber \\
   & = & \Delta \big(\langle C_{M}\rangle (I^\pi)+ \langle C_{\ell s+ \ell \ell}\rangle(I^\pi) \big) \nonumber\\
   & &+  C_r(I^\pi)+V_B(I^\pi)
   \label{MED}
\end{eqnarray}
\noindent
where the first two terms are obtained as the difference ($\Delta$) of the expectation values of $C_M$, $C_{\ell s}$ and $C_{\ell \ell}$ 
between the two mirrors. The third and forth terms correspond to the radial and ISB terms respectively.  The individual corrections and the total MED's are given in Table~\ref{coul-II}.

 \begin{table}[h]
\caption{Method 2. MED's  between $^{73}$Sr and $^{73}$Br and the contribution of each term in Eq.~\ref{MED} (in keV). }
\bigskip
\begin{tabular*}
{\linewidth}{@{\extracolsep{\fill}}|l|cccc|c|}
\hline
$I^{\pi}$&  $C_M$  & $C_{\ell s+ \ell \ell}$  & $ C_{r}$ & $V_B$  & MED\\ 
 \hline 
         $5/2^-$  &  0 & 0&  0 & 0& 0 \\
         $1/2^-$ & 11& 23&  -16& 25 & 43  \\
         $3/2^-$& -97& -130& 6 & -29 & -250 \\
   \hline
\end{tabular*}
\label{coul-II}
\end{table}

 Since the excitation energy of the $1/2^-$ state in $^{73}$Sr is not yet known we just have a lower limit for the MED of this state, which has to be greater than 27 keV. The MED value reported in Table~\ref{coul-II} is compatible with this limit but there is room for further explorations using  different values of  $\alpha_r$ for the $p_{3/2}$, $p_{1/2}$, $g_{9/2}$ and $d_{5/2}$ orbits. 
A $V_B$ contribution $\gtrsim 10$ keV in Eq.~\ref{MED} is needed to account for the MED experimental lower limit.

 \section{Conclusion}
 
 We have studied the inversion of the $I^{\pi}$= 1/2$^{-}, 5/2^{-}$ states between 
 the mirror pair $^{73}$Sr - $^{73}$Br within the framework of large-scale shell model calculations using the PFSDG-U effective interaction for the nuclear part of the problem.  The Coulomb force and other isospin-symmetry breaking effects were analyized using two well established methods which, not surprisingly, point to the prominent role played by Coulomb effects to explain the observed inversion. In Method 1 the Coulomb interaction is added to the nuclear Hamiltonian and treated both pertubatively and non-perturbatively with the calculated shifts in agreement with experiment. In this approach, possible nuclear ISB contributions might be included in the difference between neutron and proton SPE's  which are empirically derived from the spectra of $^{41}$Ca and $^{41}$Sc.   In Method 2, electromagnetic and non-Coulombic effects on the MED's are evaluated.  Within the anticipated contributions of electromagnetic origin, this second approach suggests the need for an isospin breaking nuclear contribution, 
 in line with our estimate of $V_B \approx 25$~keV, to explain the inversion.\\
\begin{acknowledgments} 
 This material is based upon work supported by the U.S. Department of Energy, Office of Science, Office of  Nuclear Physics under Contract
No.~DE-AC02-05CH11231(LBNL). AP acknowledges the support of the Ministerio de Ciencia, Innovaci\'on y Universidades (Spain), Severo
Ochoa Programme SEV-2016-0597 and grant PGC-2018-94583. AOM would like to thank the Rivas family for their hospitality during the course of this work.\\ 
\bigskip
\end{acknowledgments}

 \end{document}